# One-dimensional proximity superconductivity in the quantum Hall regime

Julien Barrier,[1,2,*] Minsoo Kim,[1,3] Roshan Krishna Kumar,[4] Na Xin,[1,2,5*] P. Kumaravadivel,[1,2] Lee Hague,[2] E. Nguyen,[1,2] A. I. Berdyugin,[1,2] Christian Moulsdale,[1,2] V. V. Enaldiev,[1,2] J. R. Prance,[6] F. H. L. Koppens,[4] R. V. Gorbachev,[1,2] K. Watanabe,[7] T. Taniguchi,[7] L. I. Glazman,[8] I.V. Grigorieva,[1,2] V. I. Fal'ko,[1,2,9] A. K. Geim[1,2,*]

[1] Department of Physics and Astronomy, University of Manchester, Manchester, UK
[2] National Graphene Institute, University of Manchester, Manchester, UK
[3] Department of Applied Physics, Kyung Hee University, South Korea
[4] ICFO — Institut de Ciencies Fotoniques, The Barcelona Institute of Science and Technology, Castelldefels, Barcelona, Spain
[5] Department of Chemistry, Zhejiang University, Hangzhou, China
[6] Department of Physics, University of Lancaster, Lancaster, UK
[7] National Institute for Materials Science, Tsukuba, Japan
[8] Department of Physics, Yale University, New Haven, USA
[9] Henry Royce Institute for Advanced Materials, University of Manchester, Manchester, UK

[*] To whom correspondence should be addressed: J.B. (julien.barrier@icfo.eu), N.X. (na.xin@zju.edu.cn) and A.K.G. (geim@manchester.ac.uk)

## Summary

**Extensive efforts have been undertaken to combine superconductivity and the quantum Hall effect so that Cooper-pair transport between superconducting electrodes in Josephson junctions is mediated by one-dimensional (1D) edge states[1–6]. This interest has been motivated by prospects of finding new physics, including topologically-protected quasiparticles[7–9], but also extends into metrology and device applications[10–13]. So far it has proven challenging to achieve detectable supercurrents through quantum Hall conductors[2,3,6]. Here we show that domain walls in minimally twisted bilayer graphene[14–18] support exceptionally robust proximity superconductivity in the quantum Hall regime, allowing Josephson junctions operational in fields close to the upper critical field of superconducting electrodes. The critical current is found to be non-oscillatory, practically unchanging over the entire range of quantizing fields, with its value being limited by the quantum conductance of ballistic strictly-1D electronic channels residing within the domain walls. The described system is unique in its ability to support Andreev bound states in high fields and offers many interesting directions for further exploration.**

## Main text

Proximity superconductivity based on quasi-1D conductors acting as weak links have attracted considerable interest from both fundamental and applied perspectives. This includes phenomena involving magnetic flux tunneling[10,19] and the associated prospect of the ampere standard based on quantum phase slips[11–13]. In terms of applications, the critical current $I_c$ in Josephson junctions (JJs) is normally suppressed by very weak perpendicular magnetic fields $B$ because of Fraunhofer-type interference between Cooper pairs propagating along different trajectories[20]. If proximity superconductivity were provided by strictly-1D states, the suppression could be avoided allowing superconducting quantum interference devices operational in high $B$. Of particular interest is the use of the quantum Hall (QH) conductors as weak links because not only this allows control of the mediating 1D states by gate voltage but also can lead to the realization of topologically-protected many-body quasiparticles (see, e.g., refs. 8,9). Despite the long-term interest in JJs incorporating QH conductors, the experimental progress has so far been limited mainly to the observation of influence of superconducting electrodes on normal-state transport and studies of so-called chiral Andreev edge states (CAES) that appear at superconductor-QH conductor interfaces[1–6,21,22]. Recently, proximity superconductivity in the QH regime was also reported for graphene-based JJs[2,6,22]. Supercurrents supported by QH edge states were found to be extremely fragile (critical current, $I_c$ of ~1 nA at mK temperatures[2,6,22,23]) so that often the proximity could not be reproduced even for devices of conceptually similar designs[20,24,25]. Below, we describe an alternative route for achieving superconducting coupling deep in the QH regime. This utilizes boundaries between AB and BA domains in Bernal-stacked bilayer graphene[14–18,26–30] which are found to serve as ballistic strictly-1D wires connecting superconducting electrodes in quantizing $B$ where the graphene bulk becomes completely insulating for Cooper pairs.

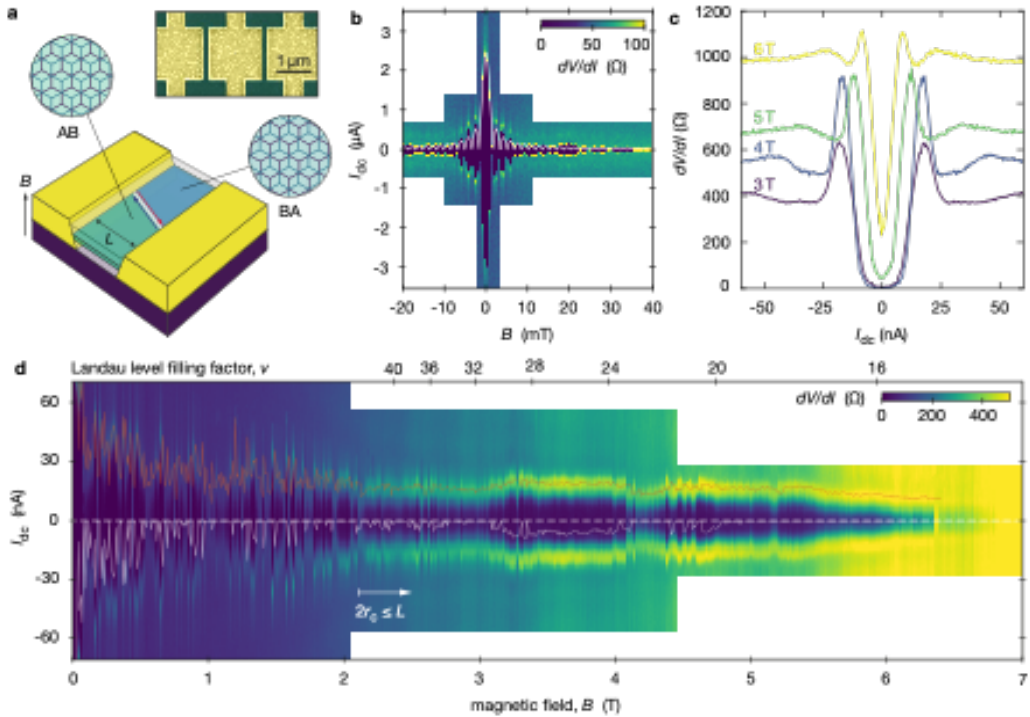

**Fig. 1| Josephson junctions incorporating domain walls in minimally twisted graphene bilayers.**
(**a**) Device schematic showing a DW acting as a weak link in the QH regime. Regions with AB and BA stacking are illustrated by the circular insets. The carrier density was varied by applying gate voltage to a Si wafer shown in dark blue (Methods). Top inset: False-color electron micrograph of a typical device containing several edgeless JJs in series. MTGB is shown in green; NbTi in yellow. (**b**) Differential resistance as a function of $I_{dc}$ in small $B$ for a junction with a single DW. Strong deviations from the Fraunhofer pattern (white curve) emerge above 10 mT. (**c**) Examples of $dV/dI$ curves in the QH regime for the same junction. (**d**) Full map measured up to 7 T in steps of 10 mT. The red curve shows $I_c$ defined as peak positions in $dV/dI$ ($I_{dc}$). The white curve marks the zero-resistance state' boundary where a finite $V$ emerged above the noise level. Data in panels **b-d** are for the same edged junction; $W \approx 3$ μm, $L \approx 200$ nm, $n \approx 2 \times 10^{12}$ cm$^{-2}$, 50 mK, $I_{ac} = 3$ nA.

The studied devices were made from minimally twisted graphene bilayers (MTGBs) as detailed in Methods. In brief, monolayer graphene was cut into two pieces that were then placed on top of each other using the parallel transfer accompanied by rotation at an angle of < 0.1° (see 'Device fabrication' in Methods). Such an assembly is known to undergo lattice reconstruction that results in formation of relatively large regions of Bernal-stacked bilayer graphene, which are separated by narrow AB/BA domain walls (DWs) with the width $w \approx 10$ nm[14,15]. The resulting domain structures could be visualized by piezo-force microscopy (Extended Data Fig. 1a) and, for MTGBs fully encapsulated in hexagonal boron nitride (see Methods), by photocurrent scanning microscopy (Extended Data Figs. 1b,c). Electron-beam lithography, dry etching and thin-film deposition were employed to make superconductor–normal metal–superconductors (SNS) junctions with MTGBs playing a role of the normal metal between superconducting (NbTi) electrodes separated by distances $L$ of ~100–200 nm (Methods). The electrodes exhibited the critical temperature $T_c \approx 7.0$ K and the upper critical field $H_{c2} \approx 9.5$ T. In total 8 devices were studied, each containing 3 to 7 MTGB junctions (Fig. 1a, Extended Data Figs. 1d,e). The junctions' width $W$ was between 0.5 and 4 μm, and they incorporated different numbers $N_{DW}$ of DWs to act as weak links between the NbTi electrodes (Fig. 1a). JJs were made in two geometries that we refer to as edged and edgeless, where graphene was either etched away everywhere, except for a narrow slit between the electrodes, or extended well beyond it, respectively (cf. schematics of Fig. 1a and Extended Data Fig. 1f). Comparison between the two geometries allowed us to assess the role played by graphene edges. As a reference, we also made similar JJs but without DWs ($N_{DW} = 0$), as well as JJs incorporating extended defects (slits and wrinkles) connecting the NbTi electrodes ('Josephson junctions without domain walls' in Methods).

In addition to the imaging, we employed normal-state electron transport to evaluate $N_{DW}$ within the examined JJs. To this end, two-probe conductance was measured at the neutrality point (NP) in high $B$ (filling factor $\nu = 0$). For JJs without DWs, their NP conductance approached zero, indicating that the MTGB bulk became insulating at $\nu = 0$ (Extended Data Fig. 2). In contrast, devices with DWs exhibited a finite zero-$\nu$ conductance with values weakly dependent on $T$ and close to $4e^2/h$ per DW, where $e$ is the electron charge and $h$ the Planck constant (Extended Data Fig. 2d). This observation agrees with the theoretical expectation that, at the NP, AB/BA walls should support chiral spin-degenerate edge states[17,18,29,30]. Good correlation was found between $N_{DW}$ estimated from our imaging and zero-$\nu$ measurements (Extended Data Fig. 2b). Because DWs could shift and even disappear from JJs during fabrication (Methods) and their number was difficult to identify from the images if DWs were close to each other, below we label JJs according to the $N_{DW}$ values found from the transport measurements.

To characterize JJs in the superconducting state, we measured their IV characteristics using small ac currents $I_{ac}$ of typically 2–5 nA and varying dc bias $I_{dc}$ ('Characterization of MTGB junctions' in Methods). First, we focus on JJs' behavior at high gate-induced electron densities (positive $n > 10^{12}$ cm$^{-2}$) which provided a low-resistance NS interface between MTGBs and NbTi electrodes (~10 Ω μm). In low $B \lesssim 50$ mT, all our devices exhibited similar characteristics, independent of $N_{DW}$ and their design (including the reference JJs). Examples are provided in Fig. 1b and Extended Data Fig. 3a that show differential resistance $dV/dI$ maps around zero $B$. They are dominated by the expected interference (Fraunhofer) oscillations, although deviations from the standard dependence (white curves) are also notable. Such behavior is typical for graphene JJs[20,25]. In the intermediate $B$ (before entering the QH regime), $I_c(B)$ did not decay $\propto 1/B$, as expected for the conventional SNS junctions, but instead exhibited giant fluctuations with numerous pockets of the zero-resistance state, which persisted up to a few T in our shortest junctions (Fig. 1d, Extended Data Fig. 4). This 'mesoscopic' behavior is characteristic of ballistic JJs[20] and, again, was observed for all our devices. Both low- and intermediate-$B$ regimes were discussed in detail previously[20,25] and are briefly reviewed in Methods. Accordingly, our emphasis below is on the proximity superconductivity that emerged in the QH regime and was exclusive to JJs containing DWs.

From the semiclassical perspective, ballistic junctions enter the QH regime if the cyclotron diameter $2r_c$ becomes smaller than $L$ so that only skipping orbits along edges (or DWs) connect the superconducting electrodes directly. In the normal state, the onset of the QH regime was evident as a rapid increase of the two-probe resistance and the concurrent appearance of Shubnikov-de Haas (SdH) oscillations (Extended Data Figs. 2a,5,10a). In this regime, no supercurrent could be discerned in any JJ without DWs, neither for the edged nor edgeless geometry, nor in reference devices (Fig. 2a, Extended Data Fig. 5), even in JJs incorporating the narrow (< 10 nm) slits that supported closely-spaced counterpropagating edge states (see Extended Data Fig. 5c and 'Josephson junctions without domain walls' in Methods). This agrees with the previous reports[20,23–25], especially taking into account our highly transparent NS interfaces such that CAES are expected to decohere at short distances[6]. In stark contrast, every JJ with DWs exhibited proximity superconductivity that extended deep into the QH regime (Figs. 1c,d; Extended Data Fig. 3b) and could approach $H_{c2}$ within ~1 T (Extended Data Fig. 4b). This shows that DWs provide an exceptionally robust channel for Cooper-pair transport. Comparing JJs with different $N_{DW}$, we found that each DW could typically carry a supercurrent of ~ 10 nA (Extended Data Figs. 3,8). To emphasize robustness and reproducibility of the DW-supported proximity, we also studied the inverse ac Josephson effect (Shapiro steps) in the QH regime and found good agreement between the experiment and theory (Extended Data Fig. 9).

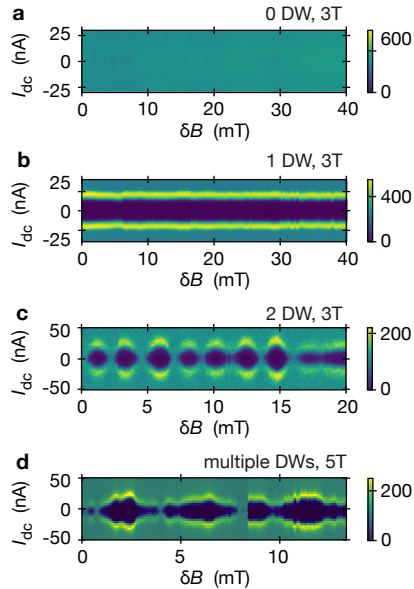

**Fig. 2| Supercurrent in the QH regime for different numbers of domain walls.** Differential resistance maps for JJs without DWs (**a**) and with one DW (**b**), two DWs (**c**) and 14 ±4 DWs (**d**). All the devices were electron doped with $n ≈ 2$-$3×10^{12}$ cm$^{-2}$. $B$ varied in steps of ~ 50 µT, $T ≈ 50$ mK, $I_{ac} = 5$ nA. Color scales in Ω. (**a-c**) measured at $B = 3$ T, (**d**) at 5 T.

Looking in more detail, for JJs with a single DW, the proximity superconductivity not only persisted deep into the QH regime but also exhibited a qualitative change in behavior such that, counterintuitively, supercurrents appeared to be stabilized by quantizing fields. Indeed, giant fluctuations in $I_c(B)$, characteristic of intermediate $B$, were suppressed for $2r_c ≲ L$ where $I_c$ remained constant over extended field intervals of ~ 0.1 T (Fig. 2b, Extended Data Figs. 6b,c). This is in contrast to the case of $2r_c > L$, where the superconductivity was confined to mT-scale pockets (Fig. 1d; ref. 20). Furthermore, $I_c$ varied relatively little over the entire interval of quantizing $B$ (despite strong and oscillating changes in the normal-state resistance) and disappeared only on approach to $H_{c2}$. On top of this gradual variation, we observed numerous abrupt changes, mostly small but occasionally substantial in magnitude (Fig. 1d,

Extended Data Fig. 7). They were irreproducible for different sweeps of $B$ and different sweep directions (Extended Data Fig. 7) and attributed to jumps of pinned vortices in the NbTi electrodes. This is generally expected because Andreev bound states responsible for Josephson coupling should depend on the superconducting order parameter in the vicinity of DWs and, hence, local vortex configurations[5,21].

For JJs with multiple DWs, the behavior could also be understood from the same perspective. For two DWs, the supercurrent was approximately twice as high as for one DW and showed oscillations nearly periodic in $B$ (Fig. 2c), as expected for interference between constant supercurrents carried by two channels. The observed periodicity in $B$ was a few times longer than that for the Fraunhofer oscillations near zero $B$, which yielded that, in the QH regime, the characteristic area per flux quantum $\phi_0 = h/2e$ was smaller than the total JJ area $L \times W$, in agreement with two supercurrent channels being present within the junction. For many DWs, the oscillating pattern became aperiodic and was interrupted more frequently by vortex jumps (Fig. 2d, Extended Data Fig. 4b). This agrees with the presence of multiple supercurrent channels, which should result in a convoluted interference pattern that is further complicated by vortices intervening at many locations.

The most revealing feature of the behavior observed for single-DW junctions is minimal variations in $I_c$ over a wide range of $B$. If the supercurrent were due to Andreev bound states arising from QH states counterpropagating at the opposite sides of the DW, one would expect Aharonov-Bohm oscillations with the periodicity $\Delta B \approx \phi_0/(w+2r_c)L < 0.1$ T where $2r_c$ accounts for the extent of QH edge states into the graphene bulk ('Steady supercurrent along a single domain wall' in Methods). No sign of such oscillatory behavior was observed in our JJs (Figs. 1d,2b; Extended Data Figs. 6,7). Even including vortex jumps, $I_c(B)$ in the QH regime varied by less than a third over intervals > 3 T (Fig. 2d), which ruled out any underlying oscillations with $\Delta B < 10$ T. The latter value translates into a spatial scale $\phi_0/\Delta BL \lesssim 1$ nm, much less than even the superconducting coherence length in NbTi. This means that QH edge states could not be responsible for the observed proximity. This is also consistent with the fact that our slits with the width of < 10 nm supported no supercurrent in the QH regime despite the nearby counterpropagating edge states. To explain $I_c(B)$ that remained steady over several Tesla, we refer to recent calculations that suggested the presence of 1D channels inside DWs[29], which differ from the well-known 1D states that appear if an energy gap is opened in the graphene bulk[17,18,26–28]. These internal channels are valley degenerate so that Andreev bound states involving the 1D electrons do not encircle any magnetic flux. This explains constant $I_c(B)$ such as shown in Fig. 2b and Extended Data Figs. 6b,c. The remaining variations in supercurrent over larger $B$ intervals can be attributed to a gradual suppression of the order parameter as vortices jump and pack up at the NS interface.

The magnitude of the supercurrents observed in the QH regime (up to 20 nA per DW) is also revealing. In zero $B$, the $I_c(T)$ dependence (Extended Data Fig. 8a) was exponential with the characteristic energy $\delta E \approx 0.2$ meV (for details, see Methods). This suggests that our ballistic JJs were in the long-junction regime where the supercurrent was limited by decoherence of Andreev bound states rather than the superconducting gap, in agreement with the previous conclusions for ballistic 2D junctions[20,31,32]. The value of $I_c$ in zero $B$ is well described by $\delta E/eR_n$ where $R_n$ is the normal-state resistance of the JJs. This is again in agreement with refs.[20,31]. It is reasonable to expect that the decoherence should be equally important for our 1D channels of the same length $L$ and, therefore, roughly the same $\delta E$ limited the critical current along DWs. This reasoning is consistent with the $T$ dependence observed in the QH regime (Extended Data Fig. 8c). Although $R_n$ for the discussed range of high $B$ and $n$ was, typically, ~ 0.5 kΩ (Fig. 2), this value arises mostly due to bulk carriers[20]. The supercurrent itself was provided by DWs and should then be limited by their resistance, that is, by $h/4e^2$ (only one 1D sub-band is expected to be occupied[29]; Extended Data Fig. 2d). Accordingly, we expect $I_c \approx (\delta E/e)/(h/4e^2) \approx 30$ nA.

This agrees well with the experiment, especially considering an additional contact resistance at the 1D-3D interface between the DW and NbTi electrodes, which should reduce $I_c$.

Finally, we discuss how the 1D proximity superconductivity was affected by the carrier density, $n$. In low $B$, the $n$-dependences were similar for all our JJs, with or without DWs (Fig. 3a). In comparison to the previous reports using JJs made from monolayer graphene[20,25], the only notable difference was the near absence of Fabry-Pérot (FP) oscillations in our devices. Such oscillations require a limited transparency of the NS interface to allow standing waves and were previously observed for hole doping where interfacial pn junctions provided suitable conditions[20,25]. The 2D-3D interface for our bilayer JJs was quite transparent even for hole doping and caused only weak FP oscillations near zero $B$ (Fig. 3a). In the QH regime, the NS interface changed its character into 1D-3D and no supercurrent could be detected for hole doping because of high resistance of the interfacial pn junctions (Fig. 3b). On the other hand, pronounced oscillations in $I_c(n)$ were observed for electron doping where the 1D-3D interface was more transparent. These oscillating are attributed to FP resonances that occur each time an integer number of half the 1D Fermi wavelength matches the domain wall length $L$ (Fig. 3b, Extended Data Fig. 10). A surprising feature of the observed FP oscillations was that their period changed little with decreasing $n$, even approaching the NP (Figs. 3b,c). This behavior described in detail in Methods seems difficult to reconcile with the fact that the electron wavelength generally diverges at zero carrier density. Nonetheless, the observed periodicity is in good quantitative agreement with that expected for our specific 1D channels (Fig. 3c) where electrons inside the DWs retain a finite density even in the case of charge-neutral bilayers[29] (see Supplementary Information).

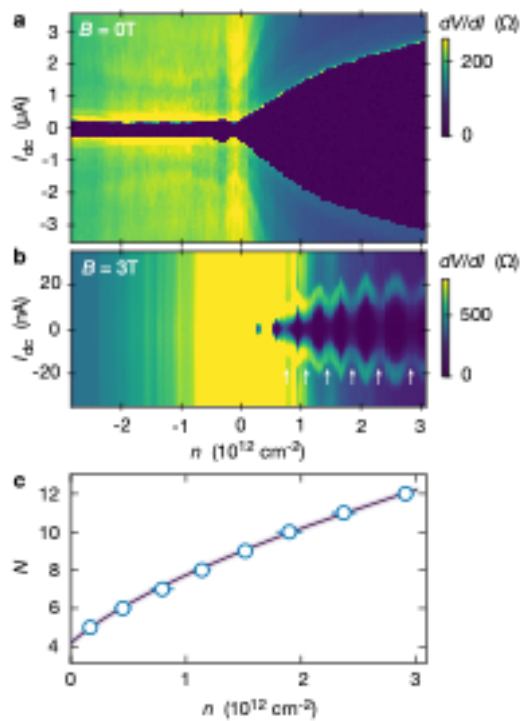

**Fig. 3| Fabry-Pérot oscillations in supercurrent provided by a single DW.** (**a** and **b**) Differential resistance as a function of doping and dc bias in zero field and the QH regime, respectively. In panel **b**, $B = 3$ T corresponds to $2r_c < L$ for all $n$. Positive and negative $n$ correspond to electron and hole doping, respectively. Same JJ as in Fig. 1, $T \approx 50$ mK, $I_{ac} = 5$ nA. White arrows in **b** indicate minima in $I_c$. (**c**) Observed minima's positions in $n$ (symbols) compared with the expected FP resonances for 1D electrons inside a 10-nm-wide DW with $L = 160$ nm (solid curve). For details, see Methods and Supplementary Information. Horizontal error bars: uncertainty is determining the minima's positions (Extended Data Fig. 10).

To conclude, AB/BA domain walls are unique in their ability to support Andreev bound states in the QH regime. The walls allow high critical currents, nearing the theoretical limit, which are practically independent of *B* due to the strictly-1D nature of electronic states inside the walls. This ballistic system offers many interesting directions for further exploration. For example, if the energy gap is opened in the bilayer graphene bulk by biasing the two layers[17], the 1D states inside AB/BA domain walls acquire topological protection[29] and should allow chiral supercurrents[3–6], which is an essential albeit not sufficient condition for the realization of non-abelian anyons[5,33]. It would also be interesting to find how the observed proximity superconductivity is affected if the spin or valley degeneracy is lifted by exchange interactions, which may, e.g., allow tunable π junctions. Furthermore, because the 1D Andreev bound states are tunnel-coupled to the graphene bulk, there is an intriguing possibility to explore interaction of the supercurrents with fractional and, especially, even-denominator QH states that were observed in encapsulated bilayer graphene and suggested to contain non-abelian quasiparticles[8]. Finally, AB/BA domain walls provide interesting venues not only within the physics of low-D superconductivity, but also in terms of normal-state transport due to their unusually long, wire-like geometry while preserving ballistic properties. Such 1D systems are exceptionally rare and could be used to address a number of phenomena in 1D including Luttinger liquids.

# Methods

## 1. Device fabrication

MTGBs were prepared using the 'cut and stack' method[35,36] with rotation by an angle of < 0.1°. Such stacks of graphene monolayers are known to form relatively large domains of bilayer graphene with the Bernal stacking order (AB and BA) which are separated by narrow (~ 10 nm) DWs[14–16,26–28,37]. After the assembly, where MTGBs were placed on top of hexagonal boron nitride (hBN) crystals, the domain structure could be visualized by piezo-force microscopy[38] as shown in Extended Data Fig. 1a. We made several JJ devices using DWs visualized by this technique. However, neither of them exhibited proximity superconductivity in the QH regime. We attribute this to further structural changes such that DWs slipped away from the proximity regions after fabrication of closely-spaced superconducting contacts. The electronic quality of the resulting JJs was also poor.

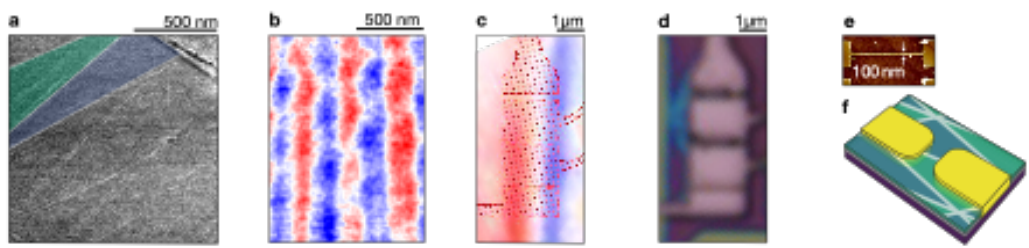

**Extended Data Fig. 1| Josephson junctions with AB/BA domain walls.** (a) Piezo-force micrograph showing domains in an MTGB before its encapsulation in hBN. The blue and green triangles indicate two neighboring regions with AB and BA stacking. (b) Photocurrent map for one of our fully encapsulated MTGB stacks that was used to make the studied JJs (photoexcitation energy of 188 meV, $n \approx 10^{12}$ cm$^{-2}$). Negative photocurrents are shown in blue, positive in red, and the white stripes in between reveal domain walls[34]. (c) Photocurrent map of a chosen DW with an overlaid design for superconducting electrodes, which is shown by the shaded red areas. (d) Optical micrograph of the same region as in panel c after depositing the electrodes. (e) Atomic-force microscopy (AFM) image of one of the studied JJs. The darker areas correspond to superconducting electrodes. (f) Schematic of our 'edgeless' devices where MTGBs extended beyond the width $W$ of JJs to avoid the presence of graphene edges in between the electrodes (compare with our 'edged' devices in Fig. 1a of the main text). The greenish triangles represent different AB and BA domains.

To preserve graphene's quality, we made MTGB structures fully encapsulated in hBN. Unfortunately, the piezo-force microscopy could not be used after an insulating hBN layer was placed on top of MTGBs[38]. To overcome this problem, we tried different methods to visualize DWs within encapsulated MTGBs and eventually converged on scanning photocurrent microscopy[34]. This dedicated technique is described in detail in ref. 34. Briefly, it utilizes scanning near-field optical microscopy (SNOM) to focus an infrared laser onto a region of interest and measures the induced photovoltage between two nearby electrodes. The resulting signal provided micrographs such as the one shown in Extended Data Figs. 1b,c where DWs appeared as blurred white stripes between red and blue regions representing neighboring AB and BA domains[34]. Note that other SNOM-based approaches were used previously to visualize DWs in twisted bilayers, revealing the characteristic triangular pattern[17,39,40]. However, for hBN-encapsulated MTGBs and in the absence of such a pattern at minimal twist angles, we found those approaches insufficient to distinguish isolated DWs from other inhomogeneities.

Using the imaged domain structures, we designed JJs trying to align DWs along the shortest distance between the superconducting electrodes (Extended Data Fig. 1c) and made devices incorporating different numbers $N_{DW}$ of domain walls. Electron-beam lithography and dry etching were then employed to embed the superconducting electrodes at the chosen positions (Extended Data Figs. 1c,d). As the superconductor, we used 60 nm of NbTi (atomic ratio: 55 to 45%) with a 3 nm thick adhesion layer of Ta. Additional 3 nm of Ta followed by 5 nm of Pt were deposited on top of NbTi to protect it from oxidation. The 4-layer film was deposited by RF sputtering at a rate of 6 nm per min under a controlled argon pressure of ~$10^5$ bar. The NbTi electrodes were found to exhibit $T_c \approx 7.0$ K and $H_{c2} \approx 9.5$ T. They were separated by the distance $L$ of 100 to 200 nm and had the width $W$ between 0.5 and 4 µm (Extended Data Fig. 1e). The devices were assembled and fabricated on top of an oxidized Si wafer that also served as a back gate to vary the carrier concentration $n$ in MTGBs.

## 2. Characterization of MTGB junctions

Electrical measurements were carried out in an *Oxford Instruments Triton* dilution refrigerator. The standard low-frequency (< 150 Hz) lock-in technique was employed using ac currents $I_{ac}$ within a few nA range. For measurements of nonlinear IV characteristics, $I_{ac}$ was superimposed on top of dc currents $I_{dc}$ ranging from nA to µA. Both ac and dc currents were sourced directly from *Zurich Instruments* lock-in amplifiers. With decreasing $I_{ac}$, differential resistance curves such as shown in, e.g., Fig. 1c of the main text and Extended Data Figs. 8b,d stopped evolving below 2 nA (i.e., they did not get sharper with decreasing $I_{ac}$), which indicated the level of electronic noise affecting our devices. The noise also limited the lowest electronic temperature ($T$) achievable for our devices to ~ 50 mK. Most measurements were done using $I_{ac}$ between 2 and 5 nA which represented a compromise between keeping $I_{ac}$ as low as possible and avoiding noise on $dV/dI$ curves, given the chosen (rather long) time constant of 1 s. Depending on the desired range for IV characteristics, $I_{dc}$ was applied in small steps $\Delta I_{dc}$ varied from < 1 nA to ~ 50 nA.

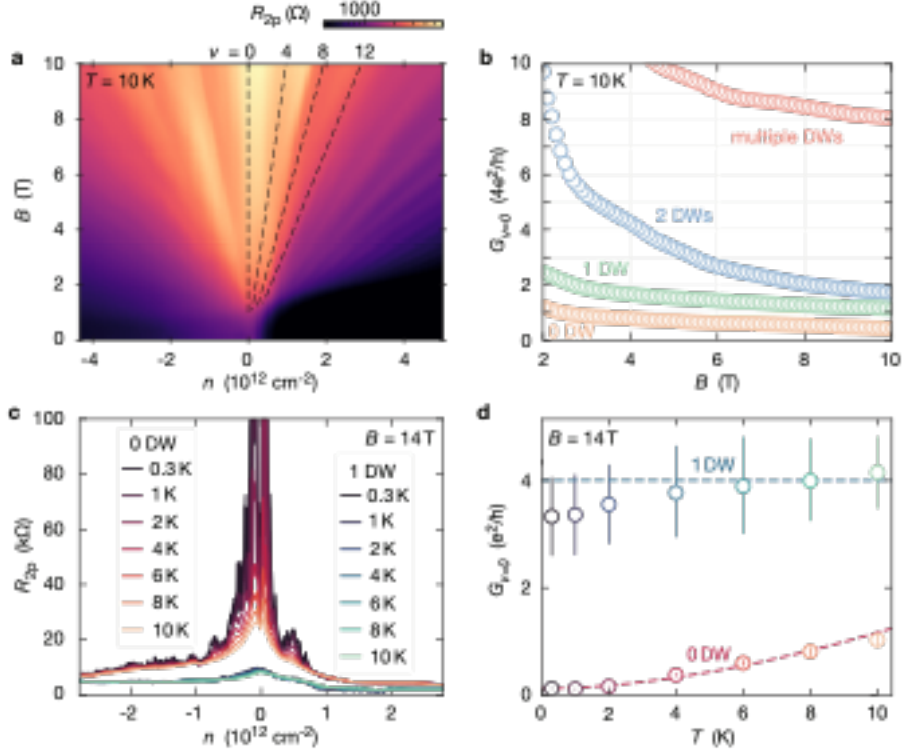

**Extended Data Fig. 2| Normal-state transport. (a)** Typical Landau fan diagram for our MTGB devices. This particular junction contained a single DW and had $L \approx 150$ nm. The filling factors ν indicated by the dashed lines were calculated using the known capacitance to the back gate; $T = 10$ K. **(b)** Two-probe conductance at the neutrality point as a function of $B$ for different $N_{DW}$. For all the plotted junctions, $L$ was between 150 and 200 nm; $T = 10$ K. **(c)** Resistance as a function of gate-induced $n$ at different $T$ for two representative junctions with 0 and 1 DWs at 14 T ($L \approx 200$ and 150 nm, respectively). Both JJs were 'edged'. **(d)** Corresponding conductance at ν = 0 (after subtracting relatively small contact resistances).

We first characterized each of the studied JJs in the normal state by measuring its two-probe resistance $R_{2p}$ as a function of $B$ and $n$ at temperatures above $T_c$, typically at 10 K. In the absence of gate voltage, all our devices were found to be slightly doped, typically by ~5×10$^{11}$ cm$^{-2}$. An example of the obtained maps $R_{2p}(n,B)$ is shown in Extended Data Fig. 2a for an MTGB junction with a single DW connecting NbTi electrodes. All the devices, independent of their design and $N_{DW}$, exhibited pronounced SdH oscillations that followed the sequence of filling factors $v = 0, 4, 8, 12, ...$, as expected for Bernal-stacked bilayer graphene. QH plateaus were neither expected[41] nor observed for this two-probe geometry.

By comparing junctions with different numbers of DWs, we noticed a clear correlation between $N_{DW}$ and $R_{2p}$ at $v = 0$ (neutrality point) such that magnetoresistance monotonically decreased with increasing $N_{DW}$. The correlations are illustrated in Extended Data Fig. 2b that plots the NP conductance $G_{v=0} = 1/R_{2p}(v = 0)$ as a function of $B$ for junctions with different $N_{DW}$. In fields above 6 T, all the two-probe curves exhibited slowly saturating $B$ dependences. For junctions without domain walls ($N_{DW} = 0$), $G_{v=0}$ saturated to small values that varied from junction to junction but were always $< 4e^2/h$ (note that $W/L \gg 1$ so that graphene's resistivity was $> 100$ k$\Omega$ per square at liquid-helium $T$), in agreement with the presence of a small gap at $v = 0$, which is expected because of both finite doping and exchange interactions[42]. Furthermore, Extended Data Fig. 2c compares devices with and without DWs at a fixed $B = 14$ T over a wider range of $T$. The latter device ($N_{DW} = 0$) exhibited a thermally activated behavior at the NP, consistent again with a small gap being present. In stark contrast, the device with a single DW showed $R_{2p}(v=0)$ that remained practically constant over the entire $T$ range (Extended Data Fig. 2c) suggesting an additional conducting channel provided by the DW.

To quantify the DW channel's conductance, we employed two complementary approaches. Using the curves such as in Extended Data Fig. 2b, we calculated the excess conductance, $\delta = G_{v=0}(N_{DW}) - G_{v=0}(N_{DW}=0)$ for JJs with DWs. The particular junction with one DW in Extended Data Fig. 2b exhibited $\delta \approx 0.8 \times 4e^2/h$ at 10 T. The junction with two DWs showed the excess conductance twice higher (within 10%) than that of the one-DW junction for all $B > 6$ T. Alternatively, assuming that the contact resistance between DWs and superconducting electrodes is close to the $R_{2p}$ value reached in the limit of high electron doping where $R_{2p}(n)$ curves saturated (Extended Data Fig. 2c), we subtracted this value as contact resistance from $R_{2p}(v = 0)$ and obtain the DW resistance itself. The corresponding $G_{v=0}$ is plotted in Extended Data Fig. 2d, which again yielded that a single DW provided the conductance of $\sim 4e^2/h$. This value is also consistent with the known electronic structure of AB/BA domain walls. Indeed, in the presence of a gap at the NP, the DWs are known to support counterpropagating (chiral) edge states which contribute the conductance quantum $e^2/h$ each and the factor of 4 comes from the spin and valley degeneracy[16,26]. Based on these observations, we used the saturation value of $G_{v=0}$ to estimate $N_{DW}$ for the studied MTGB junctions and compared it with the number of DWs seen using photocurrent scanning microscopy. Good agreement between the two values was found. The estimate for $N_{DW}$ using JJs' conductance at $v = 0$ was particularly helpful for the devices with many DWs where it was difficult to resolve individual walls by photocurrent microscopy. Further support for the described estimation was found by comparing critical currents $I_c$ in junctions with different $N_{DW}$ (next section and Extended Data Fig. 3c,d).

### 3. Supercurrents in junctions with multiple domain walls

To illustrate how the critical current behavior evolved with the number of DWs, Extended Data Fig. 3a,b shows plots for the case of large $N_{DW} \approx 15$. The plots are provided in the same representation as Figs. 1b,c of the main text for a single DW. At zero field, low $T$ and for strong electron doping $n > 10^{12}$ cm$^{-2}$, the critical current $I_c$ was of the order of a few µA per micrometer width of JJs. This zero-$B$ value did not show any systematic dependence on $N_{DW}$. In low $B$, all our JJs also exhibited pronounced deviations from the standard Fraunhofer pattern[20,25], independently of $N_{DW}$ (compare Extended Data Fig. 3a for $N_{DW} \approx 15$, Fig. 1b of the main text for $N_{DW} = 1$, and Fig. 3a of ref. 20 for $N_{DW} = 0$). Such deviations are characteristic of ballistic JJs and discussed in detail in ref. 20. In general, the observed behavior shows that the presence of AB/BA domain walls has little effect on proximity superconductivity in low $B$.

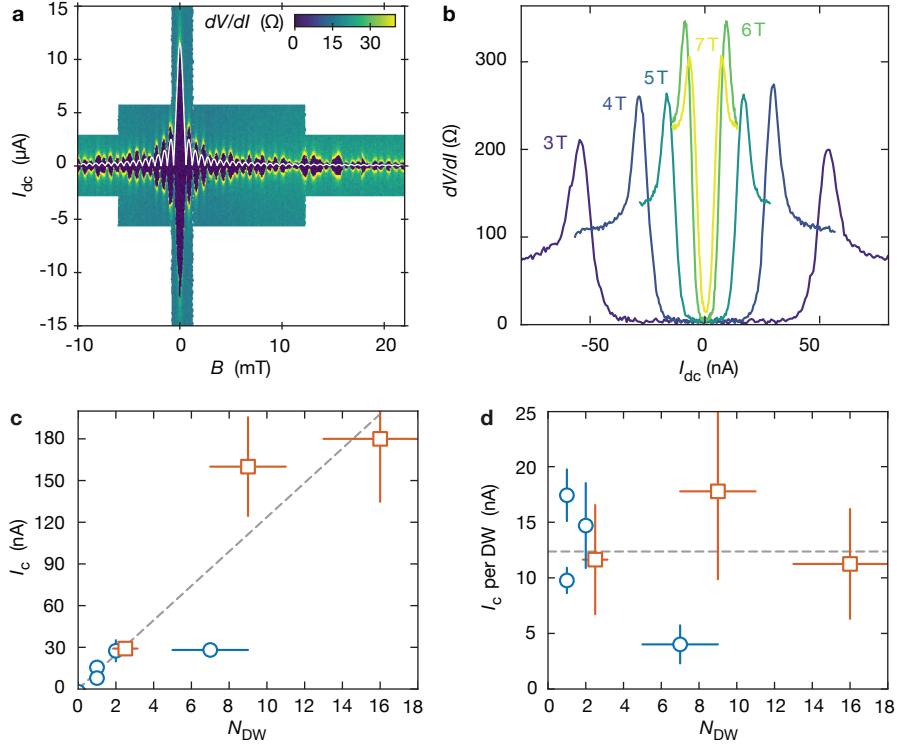

**Extended Data Fig. 3 | Supercurrent carried by AB/BA domain walls. (a)** Fraunhofer pattern typical for MTGB junctions. The shown JJ was edgeless and contained $15 \pm 3$ DWs. Measurements were done using steps in $B$ of 60 µT. White curve: standard Fraunhofer dependence $I_c(B)$ calculated using the critical current at zero $B$ and the apparent period for the first few oscillations. The deviations from the standard behavior are caused by ballistic transport of electrons and holes forming Andreev bound states[20,25]. **(b)** Differential resistance of the same junction in quantizing fields. For both (a) and (b): $T \approx 50$ mK, $n \approx 2\times10^{12}$ cm$^{-2}$, $I_{ac} = 5$ nA. **(c)** Critical current for different $N_{DW}$ ($B = 3$ T, electron doping of $\approx 3\times10^{12}$ cm$^{-2}$, $T \approx 50$ mK in all cases). Blue symbols, edged junctions; orange, edgeless ones. The dashed line is the best linear fit. The horizontal error bars are caused by uncertainty in estimating the number of DWs within the JJs. The vertical bars appear because $I_c$ rapidly fluctuated with changing $B$ and oscillated with $n$ (Extended Data Figs. 4,7; Fig. 3b of the main text) so that we plotted its rms values. **(d)** Same as in panel **c** but normalized by the number of DWs.

In the QH regime, where the cyclotron diameter $2r_c$ became smaller than the junction length $L$ so that no ballistic transport could occur through the graphene bulk[20], JJs with many DWs exhibited consistently higher $I_c$ and wider zero-resistance states than those junctions with small $N_{DW}$ (compare Extended Data Fig. 3b and Fig. 1c of the main text). Importantly, no supercurrent could be observed in the QH regime for any JJs without DWs (Extended Data Fig. 5). These observations are quantified in Extended Data Fig. 3c where $I_c$ is seen to increase roughly as $\propto N_{DW}$. This dependence suggests that each DW provided an independent Andreev channel capable of carrying a certain amount of supercurrent. Away from $H_{c2}$, the supercurrent was $\sim 10$ nA per DW at low $T$ as shown in Extended Data Fig. 3d.

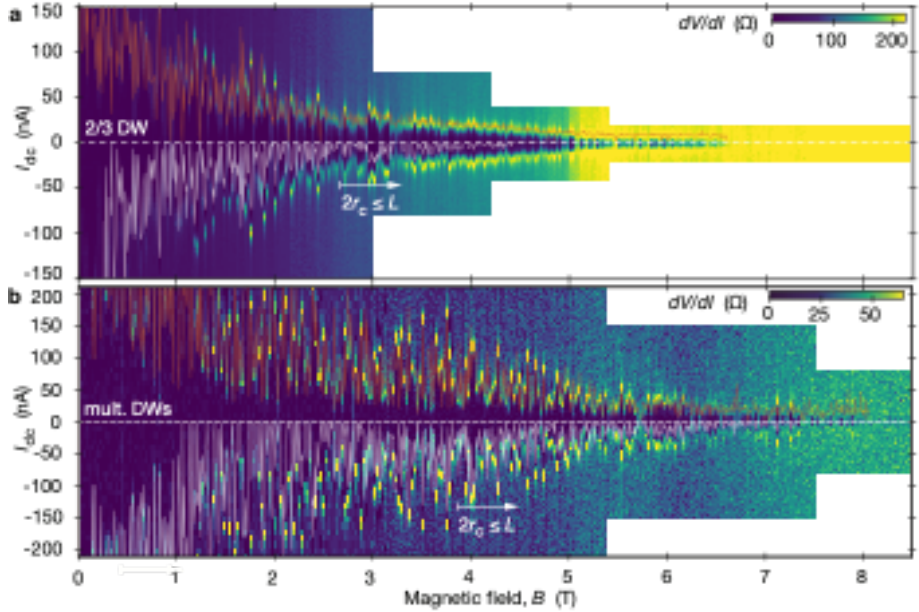

**Extended Data Fig. 4| Superconductivity in JJs with multiple domain walls.** (**a** and **b**) Differential resistance for JJs with a few (estimated as 2 or 3) and many (16 ± 3) DWs, respectively. $I_{ac}$ = 5 and 2 nA; $n \approx 2$ and $3 \times 10^{12}$ cm$^{-2}$, respectively. $T \approx 50$ mK. Both JJs were edgeless. The white curves in the bottom halves mark the boundaries of the zero-resistance state. The red curves in the top halves, the critical current. The step size in $B$ was 10 mT.

To complete the comparison between JJs with different numbers of DWs, Extended Data Fig. 4 shows differential resistance maps $dV/dI(B, I_{dc})$ over a very wide range of $B$ for junctions containing a few and many DWs. These plots should also be compared with the case of a single DW in Fig. 1d of the main text. Qualitatively, all the plots look rather similar. Supercurrent in the JJs survived in the QH regime up to fields comparable to $H_{c2}$ in the NbTi contacts, and $I_c(B)$ exhibited pronounced rapid fluctuations, independently of the number of DWs involved, if at least one DW was present (see the next section). Nonetheless, there were a couple of notable differences. First, in JJs with many DWs, finite critical currents persisted into consistently higher $B$. This is particularly obvious in Extended Data Fig. 4b where finite $I_c$ could be observed in fields reaching above 8 T, that is, within < 20% from $H_{c2}$ (compare this figure with Figs. 1c,d of the main text and Extended Data Fig. 4a). The increased $B$ range of proximity superconductivity for JJs with large $N_{DW}$ can be attributed to the simple fact that the external noise and finite $I_{ac}$, smeared our $dV/dI$ curves, so that we could detect induced superconductivity only if $I_c(B)$ exceeded a few nA. Accordingly, if many DWs contributed to the critical current, our detection threshold was breached in somewhat higher $B$. Second, in contrast to the case of a single DW, JJs with many DWs did not exhibit a clear transition from fluctuating to non-fluctuating $I_c(B)$ after entering the QH regime (compare Extended Data Fig. 4b with Fig. 1d of the main text). The strongly fluctuating $I_c(B)$ in the QH regime for the case of large $N_{DW}$ can be attributed to quantum interference between supercurrents carried by different DWs in parallel. Such interference oscillations are nearly random because many different areas are involved. The randomness is also expected to suppress the absolute value of maximum $I_c$ by a factor of 3-5 with respect to the case of 1 and 2 DWs. More importantly, vortices entering superconducting contacts in the vicinity of DWs suppress the proximity as seen on our experimental curves. This effect is much more pronounced in the multidomain devices (see, e.g., Fig. 1 and Extended Data Fig. 4), as discussed in section "Steady supercurrent along a single DW".

## 4. Josephson junctions without domain walls

To demonstrate that the robust supercurrents observed in the QH regime were due to DWs rather than any other possible mechanism[1,2,6,20,22,23,43–45], we studied JJs without DWs between superconducting electrodes (Extended Data Fig. 5). Otherwise, they were made using the same design and fabrication procedures as described above. The first type of such reference devices was based on AB-stacked bilayer graphene. These JJs were made either directly from exfoliated bilayer graphene or utilizing regions of MTGB stacks which contained no DWs (Extended Data Fig. 5a). The other reference devices incorporated either wrinkles that commonly occurred during stacking of van der Waals heterostructures (Extended Data Fig. 5b) or nanoscale slits made by high-resolution electron-beam lithography (Extended Data Fig. 5c). The general idea is that such defects in graphene can support closely-spaced counterpropagating QH edge states[1,2,6,20,22,23,43–45]. In intermediate magnetic fields ($2r_c > L$), all three types of JJs exhibited similar behavior with large fluctuations in $I_c(B)$ and interspersed pockets of the zero-resistance state (Extended Data Fig. 5). This behavior is similar to that of our JJs with DWs and, again, is attributed to ballistic transport of Andreev bound states between the superconducting electrodes[20,25]. Note that the device in panel b was edgeless, which explains the suppression of proximity superconductivity in much lower $B$ as compared with our edged JJs including those shown in panels a and c. Indeed, fluctuations in $I_c(B)$ rely on electron trajectories scattered by sample edges or extended defects[20] and are expected to be severely suppressed in edgeless JJs with parallel superconducting electrodes, in agreement with the experiment.

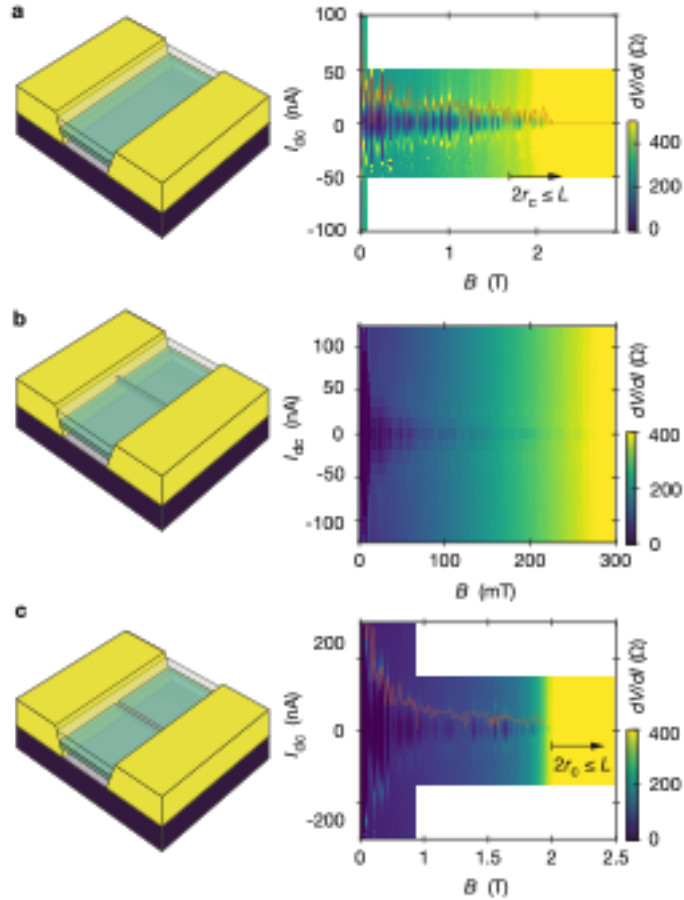

**Extended Data Fig. 5| No supercurrent in the QH regime in reference devices.** Left column, schematics of JJs. Right column, corresponding differential resistance maps at high electron doping $n \approx 3 \times 10^{12}$ cm$^{-2}$ and $L \approx 200$ nm for all the panels. Red curves, critical current. **(a)** Junction made from Bernal bilayer graphene. $W \approx 1$ μm, $I_{ac} = 5$ nA, $T \approx 50$ mK, $\Delta I_{dc} = 1$ nA. **(b)** Junction with a wrinkle formed in monolayer graphene. The wrinkle's full width was ≲ 100 nm as measured by AFM. $W \approx 1$ μm, $I_{ac} = 7$ nA, $T \approx 50$ mK, $\Delta I_{dc} = 15$ nA. **(c)** Monolayer graphene with a very narrow slit. Its width

estimated by AFM was < 10 nm. $W \approx 4$ μm, $I_{ac} = 5$ nA, $T \approx 1$ K, $\Delta I_{dc} = 1$ nA. JJ in panel **b** was edgeless; panels **a** and **c** show edged JJs.

Importantly, neither of our many reference devices exhibited any sign of proximity superconductivity in the QH regime ($2r_c < L$). Let us emphasize that no critical current in quantizing $B$ could be detected even for JJs with the narrowest slits that were less than 10 nm wide (Extended Data Fig. 5c). In this case, one can imagine Andreev states formed by QH edge states that counterpropagate along the slit's edges and are proximity-coupled through the superconducting electrodes[23]. In our slit devices, the gap in graphene was close to NbTi's coherence length $\xi \approx 6$ nm, but still no supercurrent could be discerned in high $B$. This observation agrees with recent attempts to implement the same idea using counterpropagating QH states that were located either in different graphene layers[24] or across somewhat wider ($\approx 30$ nm) slits[23]. All the evidence – from our experiments and the literature – indicates that AB/BA domain walls are unique in their ability to support Andreev bound states in quantizing $B$.

### 5. Steady supercurrent along a single domain wall

In the QH regime, JJs with multiple DWs exhibited pronounced fluctuations in $I_c(B)$ with a characteristic period of the order of one flux quantum $\phi_0$ piercing the junction area $W \times L$ (see the main text). Accordingly, these oscillations were attributed to quantum interference loops made of supercurrents propagating along different paths[2,6,22]. No oscillations with either such a short periodicity or much longer one could be observed for junctions containing a single DW (Fig. 2 of the main text). The absence of quantum interference oscillations in JJs with a single DW is reiterated by Extended Data Fig. 6. The figure shows that, similar to the device of Fig. 2b of the main text, the critical current in the QH regime was constant over rather large field intervals (Extended Data Fig. 6b,c). The junction in Extended Data Fig. 6 exhibited a monotonic decay of $I_c$ with increasing $B$, which is somewhat different from the steadier behavior for the single-DW device described in the main text (Fig. 1d). Nonetheless, the characteristic field interval $\Delta B$ over which the critical current changed considerably was at least a few T (Extended Data Fig. 6a). This again shows that any possible quantum loop made of either two supercurrent paths or counterpropagating electrons and holes forming an Andreev bound state could not be wider than $d \approx \phi_0 / \Delta B L \approx$ a few nm, that is less than $\xi$.

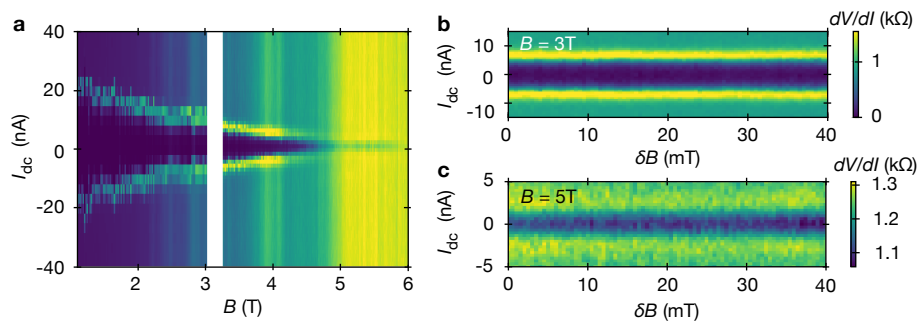

**Extended Data Fig. 6| Differential resistance maps for another junction with a single domain wall.** **(a)** Map over a large interval of $B$ (composed of two parts where the white gap indicates no data taken). Shown is an edged junction with $L \approx 150$ nm and $W \approx 0.5$ μm. Red curve, critical current. The digital noise is caused by finite steps in current: $\Delta I_{dc} = 3.3$ and $1.3$ nA below and above 3 T, respectively. Step size in $B$, 5 mT. **(b and c)** Detailed maps around 3 and 5 T, respectively. Step size in $B$, 0.5 mT. $\Delta I_{dc} = 0.6$ and 0.3 nA for panels **b** and **c**, respectively. For all the panels, $T \approx 50$ mK, $n \approx 1.7 \times 10^{12}$ cm$^{-2}$, $I_{ac} = 2$ nA. Same color scales for panels **a** and **b**.

The abrupt changes in the critical current with varying $B$, which are seen clearly in Fig. 1d of the main text, were attributed to superconducting vortices suddenly changing their positions in vicinity of the 1D-3D contacts between DWs and superconducting electrodes. To corroborate this explanation, Extended Data Fig. 7 shows two maps that were measured for the same JJ containing a single DW when sweeping the magnetic field up and down. The random nature of the jumps suggests rearrangements of vortices that were pinned within the superconducting contacts.

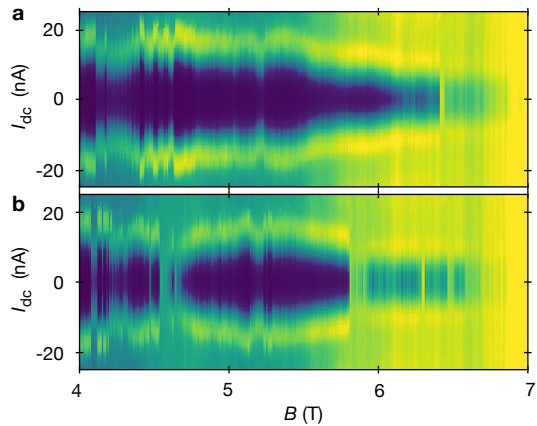

**Extended Data Fig. 7| Vortices affect the critical current in the quantum Hall regime.** Differential resistance maps for increasing **(a)** and decreasing **(b)** magnetic field in steps of 10 mT. Same device as in Fig. 1 of the main text. $T \approx 50$ mK, $n \approx 2.1 \times 10^{12}$ cm$^{-2}$, $I_{ac} = 5$ nA. Same color scale as in Fig. 1d of the main text.

### 6. Temperature dependence of the critical current

It is instructive to compare temperature dependences of $I_c$ in low and quantizing fields (Extended Data Fig. 8). In low $B$, where the proximity superconductivity is dominated by 2D Andreev-bound-state transport through the bilayer-graphene bulk, we observed a behavior similar to that reported previously for ballistic JJs made from monolayer graphene[20,31]. At $T > 2$ K, the critical current is well described by the exponential dependence[31,32]:

$$I_c(T) \propto \exp(-k_B T / \delta E) \qquad (S1)$$

where $k_B$ is the Boltzmann constant. This dependence is characteristic of so-called long JJs, in which the suppression of $I_c$ is caused by thermally-induced decoherence between energy levels of quantum-confined Andreev bound states. In ballistic junctions, $\delta E$ is expected to be $\sim hv_F/4\pi^2 L$ [31,32], which we estimate as $\sim 0.3$ meV for the device in Extended Data Fig. 8, taking into account the density dependent Fermi velocity $v_F$ in bilayer graphene but ignoring the penetration of Andreev bound states into the superconducting electrodes[32]. The latter effectively increases $L$ and makes $\delta E$ smaller. The fit in Extended Data Fig. 8a yields $\delta E \approx 0.2$ meV (white dashed curve), in good agreement with the theory estimate. This conclusion about the long-junction regime and the absolute value of $\delta E$ agrees with the previous analysis for ballistic JJs made from monolayer graphene[20,31].

At $T$ below 2 K, zero-$B$ differential resistance curves became hysteretic, exhibiting different superconducting boundaries for sweeping the dc current up and down. This is seen in Extended Data Figs. 8a,b as notable asymmetry for positive and negative $I_{dc}$. The transition between zero- and finite-resistance states happened abruptly, which resulted in seemingly diverging $dV/dI$ at the transition (Extended Data Fig. 8b). The hysteretic behavior is typical of underdamped JJs in which the switching current no longer represents the true $I_c$ (ref. 46).

In quantizing $B$, the measured $dV/dI$ curves were non-hysteretic and fully symmetric at all $T$. This is shown in Extended Data Figs. 8c,d for the case of $B = 3$ T, well above the onset of the QH regime but sufficiently below $H_{c2}$. Superficially, the temperature dependence in Extended Data Fig. 8c looks different from that in Extended Data Fig. 8a. Accordingly, it is tempting to attribute this change to a transition into the short-junction regime at high $B$, where $I_c(T)$ would no longer decrease exponentially with increasing $T$ but is expected to vary more gradually (roughly as the superconducting gap)[31,47]. The regime change also seems plausible because of the transition from 2D transport through the graphene bulk to 1D transport along DWs. However, note that $I_c$ in the QH regime at high $T$ was comparable to the probing current $I_{ac}$ (Extended Data Figs. 8c,d). Accordingly, there could be a tail of small $I_c$ extending to higher $T$, similar to the case of Extended Data Fig. 8a. Such a tail would be smeared by small but finite $I_{ac}$ and background radiation. Because of the smearing, the behavior in Extended Data Fig. 8c is inconclusive but, nonetheless, consistent with the long-junction regime, especially considering the fact that supercurrent in the QH regime (Extended Data Fig. 8c) disappeared at $T \ll T_c$ and was much smaller than in zero $B$ (Extended Data Fig. 8a).

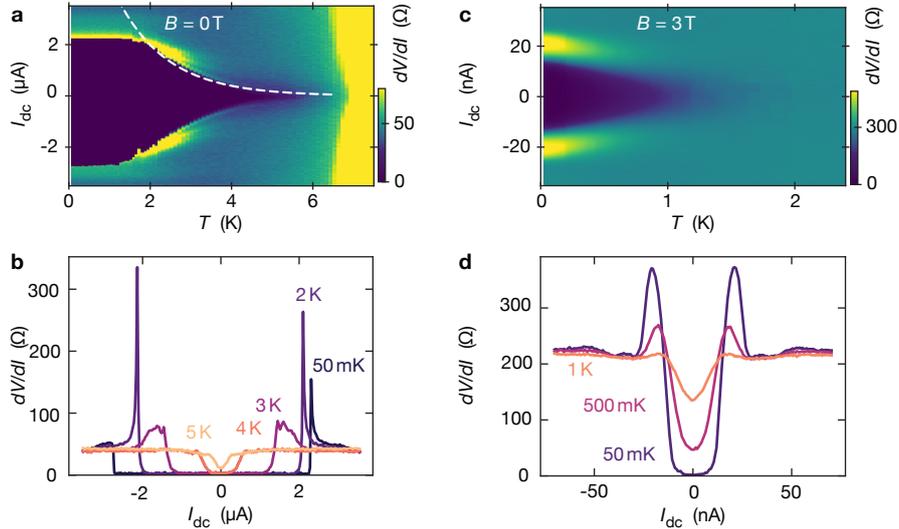

**Extended Data Fig. 8| Temperature dependence of proximity superconductivity in zero and quantizing fields.** (**a** and **c**) Differential resistance maps $dV/dI(I_{dc}, T)$ at 0 and 3 T, respectively. (**b** and **d**) Examples of $dV/dI$ for selected temperatures (cross-sections from the corresponding maps). White dashed curve in panel **a**: fit to eq. S1 above 2 K. Data are for a JJ with a single DW, $L \approx 200$ nm, $n \approx 2 \times 10^{12}$ cm$^{-2}$, $I_{ac} = 5$ nA.

## 7. Shapiro steps for 1D Josephson junctions

For completeness, we show that the proximity superconductivity along DWs in MTGBs could also be observed as the inverse ac Josephson effect. The latter effect arises from phase locking between microwave (RF) radiation and the supercurrent through JJs which leads to so-called Shapiro steps in IV characteristics. The steps appear at quantized voltages:

$$V_M = M \phi_0 f_{rf} \qquad (S2)$$

where $f_{rf}$ is the radiation frequency and $M$ the step index[48,49]. In our experiments, RF excitation was provided by signal generator R&S SMB100A and transmitted through semi-rigid coaxial cables thermally anchored to different stages of the dilution refrigerator, with attenuation of ~ 35 dB. The devices were irradiated from the cable's open end that was positioned at a distance of ≈ 1 mm from the studied JJs.

The Shapiro steps observed in the QH regime for JJs with one and multiple DWs are shown in Extended Data Fig. 9. Panel a illustrates how IV characteristics evolved as a function of the RF power $P$ at a fixed frequency. The steps gradually appeared and disappeared with varying the power, and higher order steps were clearly visible. The separation $\Delta V$ between steps increased linearly with the radiation frequency and was accurately described by eq. S2 (inset of Extended Data Fig. 9a).

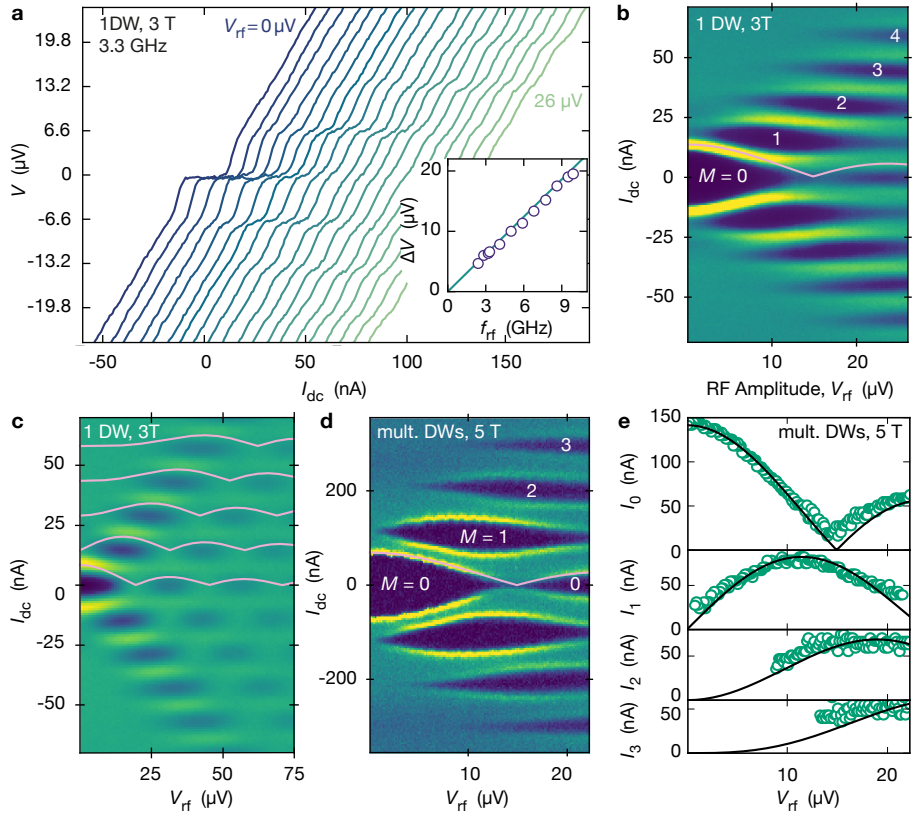

**Extended Data Fig. 9| Shapiro steps in the QH regime. (a)** Voltage vs current characteristics as a function of RF power. For clarity, the curves are shifted horizontally by 10 nA each. The power $P$ was increased in steps that corresponded to $V_{rf}$ increasing from 0 to 26 µV. Shown is the same one-DW junction as in Fig. 1 of the main text; $f_{rf}$ = 3.3 GHz, $B$ = 3 T, no $I_{ac}$ applied; $n \approx 1.8 \times 10^{12}$ cm$^{-2}$ which corresponds to a maximum in $I_c$ (Fig. 3b of the main text). Inset: $\Delta V$ as a function of the RF frequency. Green line: $\Delta V = \phi_0 f_{rf}$ as per eq. S2. **(b)** $dV/dI(I_{dc})$ with varying $V_{rf}$. The same JJ and conditions as for panel **a**; $I_{ac}$ = 5 nA. Color scale: indigo to yellow is 0 to 480 Ω. **(c)** Same as in panel **b** but for $n \approx 1.7 \times 10^{12}$ cm$^{-2}$ which corresponds to a minimum in $I_c(n)$; $f_{rf}$ = 3.52 GHz. Color scale: indigo to yellow is 70 to 440 Ω. **(d)** Similar map for a JJ with many DWs at $B$ = 5 T. $N_{DW}$ = 9±2, $L \approx$ 200 nm, $W \approx$ 3.5 µm, $n \approx 2.7 \times 10^{12}$ cm$^{-2}$, $f_{rf}$ = 3.0 GHz, $I_{ac}$ = 2 nA. Color scale: indigo to yellow is 0 to 70 Ω. **(e)** Width of Shapiro steps extracted from the map of panel **d**. The pink curves in panels **b-d** and the black curves in panel **e** are the fits by the corresponding Bessel functions as per eq. S3. For all panels, $T \approx$ 50 mK.

The width $\Delta I_M$ of the Shapiro steps is expected to follow the equation[49]:

$$\Delta I_M = |\mathcal{J}_M(V_{rf}/\Delta V)| \qquad (S3)$$

where $\mathcal{J}_M$ is the Bessel function of order $M$, and $V_{rf}$ is the ac (radiation) voltage applied along the junction. To determine $\Delta I_M$ experimentally, we measured the differential resistance $dV/dI(I_{dc})$ as a function of the RF power for fixed $B$, $n$ and $f_{rf}$ (Extended Data Figs. 9b-d). Because $V_{rf} = \alpha P^{1/2}$, we used the proportionality coefficient α as a single fitting parameter to scale the x-axes for these plots and obtain the best agreement with eq. S3[49]. The pink curves in panels b-d of Extended Data Fig. 9 show examples of the expected boundary positions for the Shapiro steps. Detailed analysis of $\Delta I_M$ for the first 4 steps is provided in Extended Data Fig. 9e. Good agreement between the experiment and eq. S3 is found for all our JJs measured under RF radiation and for both maxima and minima of supercurrent flowing along DWs (Extended Data Figs. 9b,c).

## 8. Fabry-Pérot oscillations in the critical current in the quantum Hall regime

As discussed in the main text, our JJs with AB/BA domain walls exhibited pronounced FP oscillations in the critical current. Those appeared only in the QH regime where MTGB's conductance was dominated by electron transport along DWs. An example of these oscillations was shown in Fig. 3b of the main text. Further details about the oscillatory behavior are provided in Extended Data Fig. 10. It compares the differential resistance maps $dV/dI(n,B)$ at zero and high dc biases. In the latter case ($I_{dc}$ =100 nA) and for fields above 1 T, the JJs were pushed into the normal state, in which SdH oscillations appeared (see Section 2 above; Extended Data Fig. 10a). For zero bias ($I_{dc}$ = 0), the resistance maps exhibited strong additional oscillations (Extended Data Fig. 10b). Those clearly emerged after the entry into the QH regime ($2r_c < L$) and, for the case of the JJ in Extended Data Fig. 10, persisted up to 6 T. The oscillations exhibited small changes in their $n$ positions with increasing $B$ which occurred in the direction opposite to that of SdH oscillations (Extended Data Fig. 10b). This unequivocally shows that the former oscillations were not related to Landau quantization. Note that the FP oscillations shown in Extended Data Fig. 10b do not represent oscillations in the critical current. Instead, minima and maxima in $dV/dI(n,B)$ reflect contrasting steepness of IV characteristics at different positions on the map. Nonetheless, the observed maxima in the resistance maps are expected to indicate conditions under which electron transmission through JJs was minimal and, therefore, should also correspond to minima in $I_c$. To corroborate this consideration, we measured full $dV/dI(I_{dc})$ characteristics for many fields and carrier densities, extracted the critical current values directly and plotted them as a function of both $n$ and $B$. This approach was extremely time-consuming so that we had to resort to relatively large $B$ steps of 0.5 T (Extended Data Fig. 10c). Nonetheless, FP oscillations in the critical current are clearly seen on the latter map, and minima in $I_c$ closely match maxima in the resistance oscillations of Extended Data Fig. 10b, as expected.

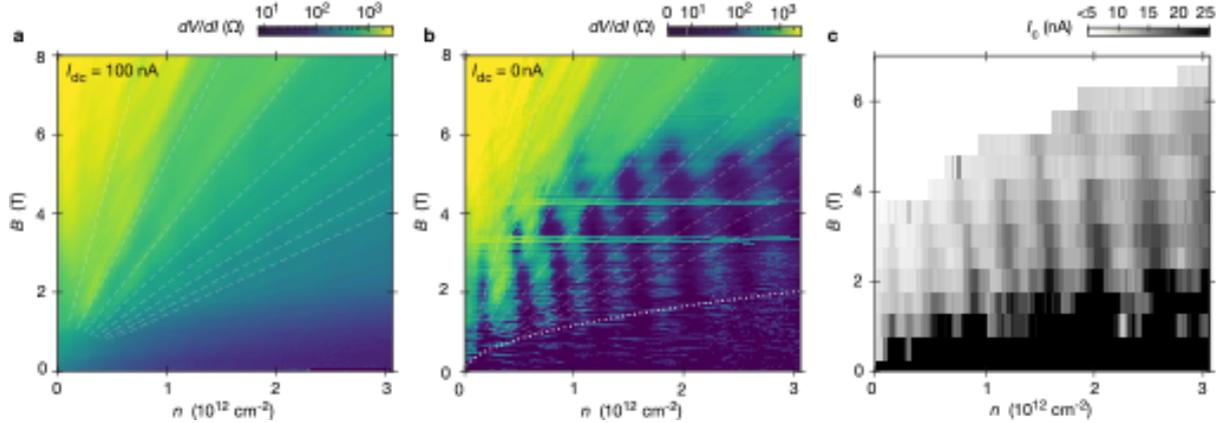

**Extended Data Fig. 10| Fabry-Pérot oscillations in the supercurrent provided by 1D states inside DWs.** (**a** and **b**) Differential resistance maps at high and low dc biases, respectively. In both cases, $I_{ac} = 5$ nA. The white dashed lines indicate the filling factors $\nu = 4, 8, 12, ...$ expected for Bernal bilayer graphene. The dotted curve in panel **b** indicates the QH regime boundary, $2r_c = L$. (**c**) Oscillations in the critical current. Values of $I_c$ are obtained from IV curves that were recorded in small steps of $\sim 3 \times 10^{10}$ cm$^{-2}$ in electron density and steps in $B$ of 0.5 T. All the measurements were carried out at $T \approx 50$ mK using JJ with a single DW and $L \approx 200$ nm.

Minima in the critical current for FP oscillations are expected to occur at integer $N = L/(\lambda_F/2)$ where $\lambda_F$ is the Fermi wavelength. Under these conditions, interference between incident and reflected electron waves within the graphene cavity between superconducting contacts leads to standing waves[20,25]. As seen in Fig. 3b of the main text and Extended Data Figs. 10b,c, the observed minima and maxima in $I_c$ occurred approximately equidistantly along the $n$-axis despite changing $n$ by more than an order of magnitude. This suggests that $\lambda_F$ of electrons responsible for the observed FP resonances changed relatively little with $n$. Such behavior cannot be explained assuming a 2D electronic spectrum, as in the case of the low-$B$ FP oscillations reported previously[20,25]. Indeed, for any 2D spectrum, $\lambda_F \propto n^{-1/2}$ which should lead to a square root dependence $N(n)$ rather than the roughly-linear one observed experimentally (Fig. 3c of the main text). To explain this surprising result, we calculated the electronic spectrum for 1D electrons confined within AB/BA domain walls and found $\lambda_F$ as a function of gate doping (Supplementary Information). The resulting curve is plotted in Fig. 3c of the main text and shows good agreement between experiment and theory. In both cases, the dependences are slightly sublinear and, importantly, do not extrapolate to zero $N$ in the limit of low densities. The latter observation reflects the fact that AB/BA domain walls support a finite electron density within charge-neutral MTGBs[29]. The observed small shift of the FP resonances towards lower $n$ with increasing $B$ remains to be understood (Extended Data Figs. 10b,c). Tentatively, we attribute the shift to field-induced changes in an electrostatic confinement of 1D electrons, which are not accounted for in the model described in ref. 29.

## 9. One-dimensional electrons in AB/BA domain walls

Although the theory provided in this supplementary section was presented in ref. 29, we believe it would be useful to briefly review the calculations and somewhat simplify them, focusing on parameters most suitable for our experimental situation and, in particular, the case of DWs with $w \approx 10$ nm[14,18,50]. As a model, we consider a partial screw dislocation aligned along the armchair direction, that is the most energy favorable direction for AB/BA domain walls in MTGBs[18]. Local deformations around such a DW are illustrated in Supplementary Fig. 1 where the color scale indicates a relative shift between top and bottom graphene lattices. The dislocation's Burgers vector is $(0, a/\sqrt{3})$, where $x$ and $y$ axes denote zigzag and armchair directions, respectively. The corresponding displacements can be found by solving the Frenkel-Kontorova model[51] and are given by[29,50]:

$$\vec{u}(x) \approx \left(0, \frac{2a}{\pi\sqrt{3}} \arctan(e^{2x/w})\right),$$

where the interlayer shift is assumed to be zero for AB stacking (large negative $x$) and reaches the Burgers vector value on the BA side. The effective width $w$ of AB/BA walls depends on the interlayer adhesion energy and graphene's elastic constants but may also be affected by the presence of encapsulating hBN, DW orientation and strain induced during fabrication. Below we assume $w = 10$ nm, as estimated in ref. 50 and consistent with experimental images of DWs in MTGBs[14,18].

To describe electronic states in the vicinity of DWs, we use a $4 \times 4$ hybrid $k \cdot p$ – tight-binding Hamiltonian[29,51,52]:

$$\widehat{H}_\pm = \begin{pmatrix} \hbar v \vec{\sigma}_\pm \cdot \vec{k} + \frac{1}{2}\Delta & \widehat{T}_\pm \\ \widehat{T}_\pm^\dagger & \hbar v \vec{\sigma}_\pm \cdot \vec{k} - \frac{1}{2}\Delta \end{pmatrix},$$

with a position-dependent (local) interlayer hopping:

$$\widehat{T}_\pm = \frac{\gamma_1}{3} \sum_{j=0,1,2} e^{\pm i(\vec{K}_j - \vec{K}_0)\cdot \vec{r}_0} \begin{pmatrix} 1 & e^{\pm ij2\pi/3} \\ e^{\mp ij2\pi/3} & 1 \end{pmatrix} = \frac{\gamma}{3}\begin{pmatrix} \theta_1 & \theta_2 \\ \theta_2 + \theta_3 & \theta_1 \end{pmatrix}$$

$$\theta_1 = 1 + 2\cos\left(\frac{2\pi}{3} + \varphi\right); \quad \theta_2 = 1 + 2\cos\varphi; \quad \theta_3 = 2\sqrt{3}\sin\left(\varphi - \frac{\pi}{3}\right); \quad \varphi = \frac{2\pi}{3}\frac{\sqrt{3}u(x)}{a}.$$

Here, $\vec{\sigma}_\pm = (\pm\sigma_x, \sigma_y)$ is the 'vector' of two Pauli matrices for $\pm K$ valleys, $\gamma_1 \approx 380$ meV is the interlayer hopping parameter, $\vec{r}_0 = \left(0, \frac{1}{\sqrt{3}}a + u\right)$ is the lateral interlayer offset counted from AA stacking and $\pm\vec{K}_{0,1,2}$ are the triads of equivalent Brillion zone corners corresponding to $\pm K$ valleys. Note that hopping matrices $\widehat{T}_\pm$ are same in the two valleys, and $\vec{k} = -i(\partial_x, \partial_y) - \frac{e}{\hbar}(0, Bx)$ incorporates the magnetic vector potential (in the Landau gauge).

Without magnetic field, the calculated spectra in the vicinity of a DW are shown in Extended Data Fig. 11b. The black dots represent energies $\varepsilon(k_y)$ of 1D electronic states propagating along the DW. They lie below the nearly-parabolic continuum arising from the bulk bilayer graphene (as indicated in blue) and are clearly separated from it. Upon merging into the continuum, the 1D states become quasi-stationary, that is, they mix with bulk states and acquire a finite lifetime, so that we cannot identify the quasi-stationary states among those computed at high $k_y$ values at zero $B$ (in the blue region of Extended Data Fig. 11b).

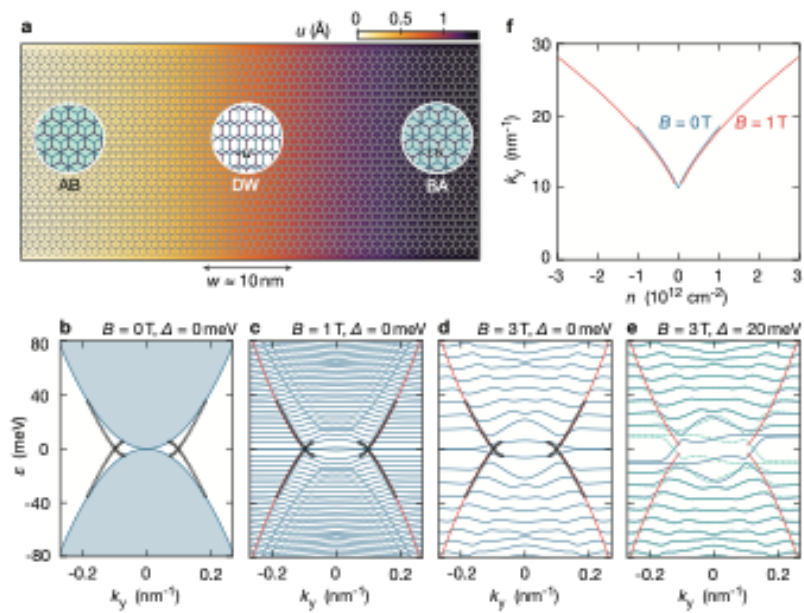

**Extended Data Fig. 11| Electronic spectra of AB/BA domain walls. (a)** Local distortions associated with DWs lying in the armchair direction, which were used in our calculations of the electronic spectra[29]. The insets illustrate relative positions of atoms in top and bottom graphene layers. **(b)** Electronic states associated with AB/BA domain walls (black dots). The blue parabolas denote the bulk spectrum, away from the DW. **(c-e)** Calculated spectra in finite **b** (blue curves). $B = 1$ T for panel **c** and 3 T for panels **d** and **e**. A gap of 20 meV is opened in the bulk spectrum of panel **e**. The flat parts correspond to non-dispersive LL states in bulk bilayer graphene. The orange curves connect points of the steepest dispersion and correspond to mixing of LLs with 1D states, which gives rise to the dispersive features. The 1D states are centered at zero $k_y$, which corresponds to the DW's center, $X = \hbar k_y/eB = 0$. The black-dot curves in panels **c** and **d** are taken from zero-$B$ calculations to show that the features originate from the same 1D DW-bound states[29]. In the absence of interlayer bias, $\pm K$ valleys are represented by the same (blue) curves. The valley degeneracy is lifted in panel **E** by interlayer bias, where the solid blue and dotted green curves represent $\pm K$ valleys. **(f)** Dependence of $k_y^F(n)$ for the 1D states. The blue curve was found using $\varepsilon(k_y)$ at zero $B$ and the orange curve using the dispersion represented by the orange solid curve in panel **c**.

In quantizing $B$, the spectral continuum splits into Landau levels (LLs) shown in Extended Data Figs. 11c,d. The levels are nondispersive sufficiently away from the DW with wavefunctions being centered at distances $X = \hbar k_y/eB$ from the middle line $X = 0$. Several sets of dispersive features can also be seen in the figures. Some of them evolve with increasing $B$, indicating that they originate from skipping orbits (QH edge states) induced by DWs. However, there is a special set of dispersive features, which closely follows the 1D states computed at $B = 0$ and remains practically unaffected by magnetic field (Extended Data Figs. 1c,d). These states are marked by orange lines drawn through the points of highest drift velocities $\partial \varepsilon/\partial k_y$ found for different LLs. This spectral behavior reveals that quasi-stationary 1D states associated with the DW get stabilized by the opening of cyclotron gaps and also become mixed perturbatively with LLs, giving the latter the dispersive features highlighted in orange in Supplementary Fig. 1. Despite the mixing, the 1D states are expected to propagate long distances along DWs before they get affected by the bulk[29]. The gap induced by interlayer bias (due to doping by gate voltage) is found to have little effect on the DW-bound states at higher energies (Extended Data Fig. 11e). Importantly, the 1D states are centered at $X = 0$ for both valleys and, within each valley, can propagate in both directions along the same trajectory. This means that these DW-bound states are nonchiral as they have the same dispersions for both $\pm K$ valleys, in contrast to the well-known valley-polarized chiral modes[14–18,26–28] which appear inside the BLG gap upon symmetry breaking by interlayer bias (Extended Data Fig. 11e).

Using the described calculations, we evaluated the Fermi wavevector $k_y^F$ for 1D DW-bound states as a function of the global carrier density in bilayer graphene[29]. To this end, we used the known dependence for the Fermi energy $E_F$ on $n$ for the bulk bilayer and then found $k_y^F$ from the dependences $\varepsilon(k_y^F) = E_F$ calculated in Extended Data Figs. 11b,c. The resulting $k_y^F(n)$ is plotted in Extended Data Fig. 11f and was used to fit positions of the FP resonances found experimentally (see Fig. 3c of the main text). Let us emphasize that the finite value $k_y^F \approx 0.1 \, \text{nm}^{-1}$ computed for a neutral MTGB reflects the fact that the 1D dispersion in Extended Data Fig. 11b crosses the conduction-valence band edge of the bulk dispersion in bilayer graphene.

## Note added in proof

After the manuscript was accepted, a series of papers was brought to our attention, including refs. 53 and 54. They report proximity superconductivity in 3D bismuth nanowires in high magnetic fields and attribute it to chiral surface states. Although not directly related to the subject of our report (superconductivity in the quantum Hall regime), the work can be of interest for experts working on topologically-protected quasiparticles and Josephson junctions in general.

## Methods' references

## Acknowledgements

We acknowledge financial support from the European Research Council (grant VANDER), the Lloyd's Register Foundation, Horizon 2020 Graphene Flagship Core3 Project, and the EPSRC (grants EP/V007033/1 and EP/S030719/1). J.B. acknowledges support from the EPSRC (Doctoral Prize Fellowship). R.K.K. acknowledges the EU Horizon 2020 program under grants 754510, 893030 and the FLAG-ERA program (PhotoTBG). L.I.G. acknowledges support from the National Science Foundation through Grant No. DMR-2002275, and the Office of Naval Research (ONR) under award number N00014-22-1-2764.

## Author contributions

A.K.G. and J.B. initiated and led the project. N.X. and P.K. fabricated the devices with help from L.H. Domain walls in MTGBs were imaged by R.K.K., F.H.L.K. and R.V.G. J.B. carried out the electrical measurements with help from M.K., E.N., A.I.B., and J.R.P. J.B. and A.K.G. analyzed data with help from I.V.G., L.I.G., J.R.P., and V.I.F. C.M., V.V.E., L.I.G. and V.I.F. provided theoretical support. K.W. and T.T. supplied quality hBN crystals. J.B., I.V.G. and A.K.G. wrote the manuscript with contributions from N.X. and V.I.F. All authors contributed to discussions.

Correspondence and request for materials should be addressed to Julien Barrier (julien.barrier@icfo.eu), Na Xin (na.xin@zju.edu.cn) or A. K. Geim (geim@manchester.ac.uk).

## Data availability

Original data files that support the findings of this study are available on doi:10.5281/zenodo.10698874 and from J. Barrier.